# Doping single wall carbon nanotubes with differently charged porphyrins.

G. A. M. Sáfar, H. B. Ribeiro, F. O. Plentz.
*Departamento de Física, Universidade Federal de Minas Gerais, 30123-970, Belo Horizonte, Brazil*
C. Fantini, A. P. Santos
*Centro de Desenvolvimento da Tecnologia Nuclear,CDTN/CNEN, Belo Horizonte-MG, 30123-970, Brazil*
G. DeFreitas-Silva, Y. M. Idemori
*Departamento de Química, Universidade Federal de Minas Gerais, Belo Horizonte-MG, 30123-970, Brazil.*

Photoluminescence (PL) measurements of porphyrin-doped single wall carbon nanotubes (SWNT) were studied in sodium dodecylbenzenesulfonate (NaDDBS) aqueous dispersions. The PL spectra were used to draw PL maps were the maxima corresponds to absorption-emission excitonic processes related to (E11, E22) first Van Hove singularities of the SWNT electronic structure. The influence of the net charge of the porphyrin was a determinant factor in the energy map maximum shifts (EMMS) compared to the energy map of a pristine NaDDBS/SWNT dispersion. A non-interacting porphyrin is used as a reference to discard the influence of the dielectric constant of the medium in the EMMS.

The investigation in the interactions between porphyrins and carbon nanotubes has been raised since the first study on the subject appeared and since the elaboration of photoelectrochemical cells that uses these two materials as building blocks began [1, 2]. Besides the technological concern on the subject, the underlying phenomena related to the photoinduced charge transfer are of general interest in light harvesting by aromatic compounds that changes energy by proximity interactions to allotropic forms of carbon.

Since the early discovery of single wall carbon nanotubes (SWNT) by Iijima, photoluminescence studies in these materials have been carried out [3, 4, 5].

Recently, optical measurements of porphyrins/SWNT were used to study the porphyrin/carbon nanotubes interactions. Most studies were concerned in the influence of SWNTs in porphyrin fluorescence rather than the influence of porphyrin in SWNT photoluminescence (PL) emission [1, 6]. Even more recently, PL spectroscopy in the near infrared, the energy range of PL emissions of small diameter SWNT, was used to study these charge interactions [7]. In a previous work, we were able to determine the influence of molecular doping of SWNT by porphyrin by PL energy maps [8].

In this work, we investigate the PL absorption-emission processes of SWNT/porphyrins close interactions in surfactant/aqueous media.

Experimental

The experimental procedure for preparing the samples is described in a previous work [8]. Briefly, HiPco SWNT were dispersed in sodium dodecylbenzenesulfonate (NaDDBS) aqueous solutions by high power sonication. After, the porphyrin aqueous solutions were mixed with the SWNT dispersions and sonicated. The porphyrins used were meso-tetrakis(4-carboxyphenyl)porphyrin, or shortly, $H_2TCPP$, and meso-tetrakis(4-sulfonatophenyl)porphyrin (sodium salt), or shortly, $H_2TSPP$.

HiPco SWNT were dispersed in aqueous solution (concentration 0.05 mg/ml, pH 7.4) with NaDDBS 1 %wt. in deionized water. The dispersion was made by sonicating the solution for 30 min in 10W ultrasonic tip and then centrifuged at 28400 g by 90 min.

PL spectra in the NIR range taken from the supernatant were recorded. Afterwards, the sample was mixed with a solution with each porphyrin and sonicated in a Fisher scientific FS220 ultrasonic bath for 30 min. After, PL spectra of the resulting sample were measured in the UV-visible range with a 457 nm excitation pumped by an $Ar^+$ laser and absorption spectra in UV-vis (190-800 nm) were recorded in an HP-8453A diode-array spectrophotometer. PL spectra in the near IR range were recorded again to plot the PL maps. All the optical measurements were recorded at room temperature.

For the PL maps, the sample was excited with different excitation energies using a Ti:sapphire laser, pumped by an $Ar^+$ laser. Photoluminescence spectra were recorded in a backscattering geometry and the output signal was focused onto an InGaAs diode array detection system and a Spex 750M Spectrometer through a microscope objective.

Results

A linear interpolation routine in MatLab was used to construct the contour plot for both samples and it is shown in Fig. 1. The maxima of the PL maps were identified and are discriminated in Fig. 2. The results from a previous work are listed for comparison along with the results from this work.

The SWNT/NaDDBS/$H_2$TCPP dispersion shows a red-shift in the energy maxima compared to the SWNT/NaDDBS dispersion, both in emission and absorption, for all nanotube (n,m) indices. The SWNT/NaDDBS/$H_2$TSPP dispersion shows smaller energy shifts, for upper or lower energies, depending on the nanotube (n,m) index. We plot the maxima reported here along with the ones obtained in ref [8] in Fig 2. There, the maxima obtained are blue-shifted to the ones of the SWNT/NaDDBS dispersion.

The optical absorption spectra of the SWNT/NaDDBS/$H_2$TCPP dispersion show a distorted Soret band, as the ones shown by SWNT/NaDDBS/$H_2$TTMAPP, in ref [8]. However, the Soret band in the SWNT/NaDDBS/$H_2$TSPP dispersion is quite the same shown by the $H_2$TSPP solution (data not shown).

Discussion

The absence of distortion of the Soret band in the SWNT/NaDDBS/H$_2$TSPP dispersion can be related to a lack of charge transfer between the carbon nanotubes and the H$_2$TSPP ions in solution. This could also explain the little variation of the PL maxima of that dispersion compared to the SWNT/NaDDBS dispersion. One could suppose that the surfactant barrier of NaDDBS cannot be pierced by the sonication. However, the other dispersions were submitted to the same conditions.

Nevertheless, in solution, the net charge of H$_2$TSPP is 4-, the same one of H$_2$TCPP. Concurrently, the acid strength of both, as their molecular polarity, is similar. Therefore, the dielectric constant of an aqueous solution containing either one of the molecules is quite similar. The difference between H$_2$TSPP and H$_2$TCPP is that hydrogen bonds can be formed between molecules of H$_2$TCPP, due to the carboxyl groups. This facilitates the aggregation of molecules, either side-by-side or stacked [9].

In ref [8], we attributed the energy shift of porphyrin/SWNT/NaDDBS dispersion from SWNT/NaDDBS dispersion to doping and to a dielectric constant variation. In that work, we suggested that H$_2$TTMAPP, another porphyrin, with molecular net charge 4+, in SWNT/NaDDBS/H$_2$TTMAPP dispersion, was doping the SWNTs, because the most recent theoretical analyses predict a red-shift for a dielectric constant variation [10, 11]. Fig. 2 shows that the energy maxima of the PL map of SWNT/NaDDBS/H$_2$TTMAPP are blue-shifted compared to the ones of SWNT/NaDDBS dispersion.

We suggest that doping is responsible for the red-shift of SWNT/NaDDBS/H$_2$TCPP and the blue-shift of SWNT/NaDDBS/H$_2$TTMAPP, with both shifts measured from the pristine SWNT/NaDDBS dispersion. The reason is that a little or absent interaction of H$_2$TSPP with the SWNT in the dispersion has a minor effect on the energy maxima of the PL. If the dielectric constant of the environment is similar for both H$_2$TSPP and H$_2$TCPP, the red-shift caused by H$_2$TCPP is mainly caused by charge transfer between H$_2$TCPP and the SWNTs. The H$_2$TCPP Soret band distortion caused by the SWNTs corroborates it.

The difference between the energy shifts for each SWNT index could be due to different aggregation states of the porphyrin molecules on the SWNT surface [9]. In that case, further analysis is challenged by the presence of the NaDDBS in the SWNT surface.

Conclusion

Porphyrin molecules are able to dope SWNT according to their molecular net charge. A blue-shift in the energy maxima of the PL maps occurs when a positively charged porphyrin (H$_2$TTMAPP) is added to a SWNT/NaDDBS dispersion compared to the pristine dispersion. Instead, a red-shift appears when a negatively charged porphyrin (H$_2$TCPP) is added to the same pristine dispersion. As a general trend, little or none energy shift appears when there is no charge transfer between porphyrin (H$_2$TSPP) and SWNT. Combining both facts, we suggest that the dielectric constant variation has a minor effect in energy shifts of the PL map maxima compared to molecular doping of the SWNT/NaDDBS aqueous dispersion by the porphyrins.

Acknowledgements

We thank Marcos Assunção Pimenta for useful discussions, and CNPq and FAPEMIG for the financial support.

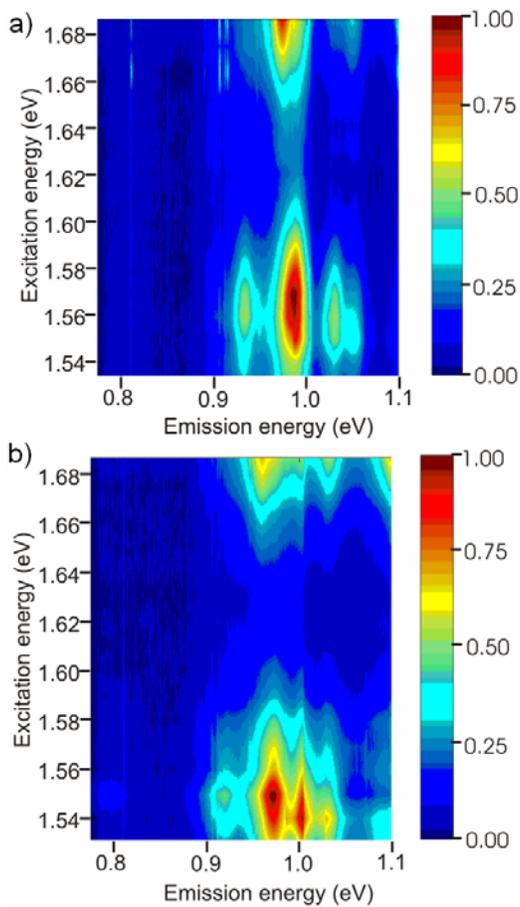

Figure 1: a) PL map of H$_2$TSPP/SWNT/NaDDBS and b) PL map of H$_2$TTMAPP/SWNT/NaDDBS. The color scale is linear and normalized.

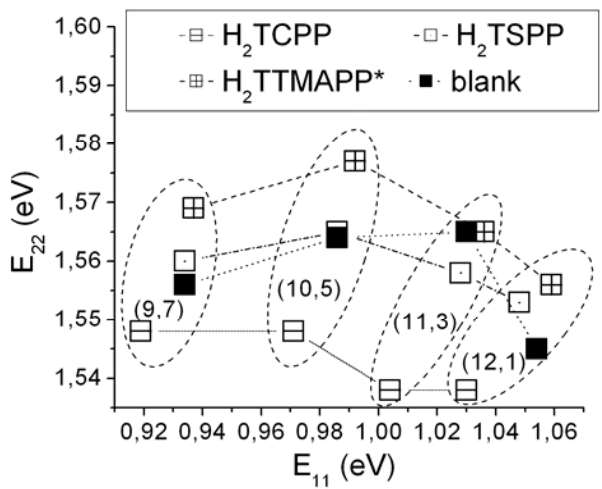

Figure 2: PL map maxima of the dispersed porphyrin/SWNT/NaDDBS samples. The legend indicates which porphyrin is added. Blank means no porphyrin added. The straight lines connect the same sample maxima. The ellipsis encircle the same nanotube (n,m) index. Asterisk refers to ref. [8].